\documentstyle[prb,aps]{revtex}
\begin{document}
\draft\input{psfig}
\title{A complete devil's staircase in the Falicov-Kimball model}
\author{C. Micheletti$\null^1$, A. B. Harris$\null^2$ and
J. M. Yeomans$\null^1$}
\address{(1) Theoretical Physics, Oxford University,
1 Keble Road, Oxford OX1 3NP, UK}
\address{(2) Department of Physics,
University of Pennsylvania, Philadelphia, PA 19104-6396}
\date{June 1996}
\maketitle
\begin{abstract}
We consider the neutral, one-dimensional Falicov-Kimball
model at zero temperature in the limit of
a large electron--ion attractive potential, $U$. By calculating the
general $n$-ion interaction terms to leading order in $1/U$ we argue
that the ground-state of the model exhibits the behavior of a
complete devil's staircase.
\end{abstract}


In this letter we study the ground-state phase diagram of the
one-dimensional Falicov-Kimball model. This model was proposed to
describe metal-insulator transitions [\onlinecite{FK}] and has since
been investigated in connection with a variety of problems such as
binary alloys [\onlinecite{FF}], ordering in mixed-valence systems
[\onlinecite{RF}], and the formation of ionic crystals
[\onlinecite{GLM}]. It is the latter language we shall use here,
considering a system of static positive ions and mobile spinless
electrons. The model comprises no electron-electron or ion-ion
interactions but an on-site electron-ion attraction, $-U$.

We write the Falicov-Kimball model in the form
\begin{eqnarray}
&&{\cal H}= t \sum_j (a^\dagger_j a_{j+1} + a^\dagger_{j+1} a_j) - U
\sum_i (s_j a^\dagger_j a_j - 1/2) \nonumber \\
&&+ (U/2 - \mu_i) \sum_j (s_j - 1/2) +
(U/2 - \mu_e) \sum_j (a^\dagger_j a_j - 1/2)
\label{eqn:ham}
\end{eqnarray}

\noindent where $a^\dagger_i$ ($a_i$) denotes the fermionic creation
(destruction)
operator for a spinless electron, $s_i$ is equal to 1 (0) if site
$i$ is (un)occupied by an ion, $t$ is the hopping integral for
electrons,  $\mu_i$ and $\mu_e$ are the chemical
potentials for ions and electrons respectively and $U$ is
a positive constant corresponding to the ion--electron attractive
energy. The choice of a positive $U$ is not restrictive since the
transformation $\{ U \to -U ; \mu_i \to - \mu_i; s_i \to 1 - s_i
\}$ maps the Hamiltonian (\ref{eqn:ham}) onto the same system with $U$
negative.

 The ground state of the system is
chosen by minimizing the energy per site over all possible ionic
arrangements. The structure of the ground states differs significantly
depending on whether $U$ is large or small compared to $t$. In the
first case the electrons are essentially localized near the ions and
the latter tend to be as far apart as possible while, for large $t/U$,
the delocalization of electrons favors the formation of clusters of ions
[\onlinecite{L}]. In this letter we consider
the case where $U$ is very large compared to other parameters in
(\ref{eqn:ham}), and treat $t/U$ as a perturbative parameter.

Despite the simplicity of the Falicov-Kimball model the determination
of the ground
state is far from trivial. Numerical results [\onlinecite{GUJ}] have
suggested that in
the neutral system, where the number of electrons and ions are equal, a
large number of modulated phases appear as ground states.
In 1989 Barma and Subrahmanyam studied the phase diagram of the model
by mapping it onto an Ising system [\onlinecite{BS}]. They showed that
the phases appearing at the first few stages of a perturbative
analysis could be described in terms of a simple branching rule, hence
suggesting that the complete phase diagram
might display a devil's staircase. A different approach to
the large-$U$ limit was later introduced by Gruber {\em et
al.}\/ [~\onlinecite{GLM}~] who considered the model as a set of ions
with interactions mediated by
the electrons. They calculated the two-ion interaction to leading
order in $t/U$ on the basis of which they argued that the ion spacing
is constant in the ground state.

Here we show that a full determination of the ground state requires a
calculation of the general $m$-ion interactions. These are obtained to
leading order in $t/U$ using Green's function techniques. Then, using
arguments first introduced by Fisher and Szpilka [\onlinecite{FS}], we
deduce the existence of a devil's staircase in the neutral
Falicov-Kimball model.

\begin{figure}
\centerline{\psfig{figure=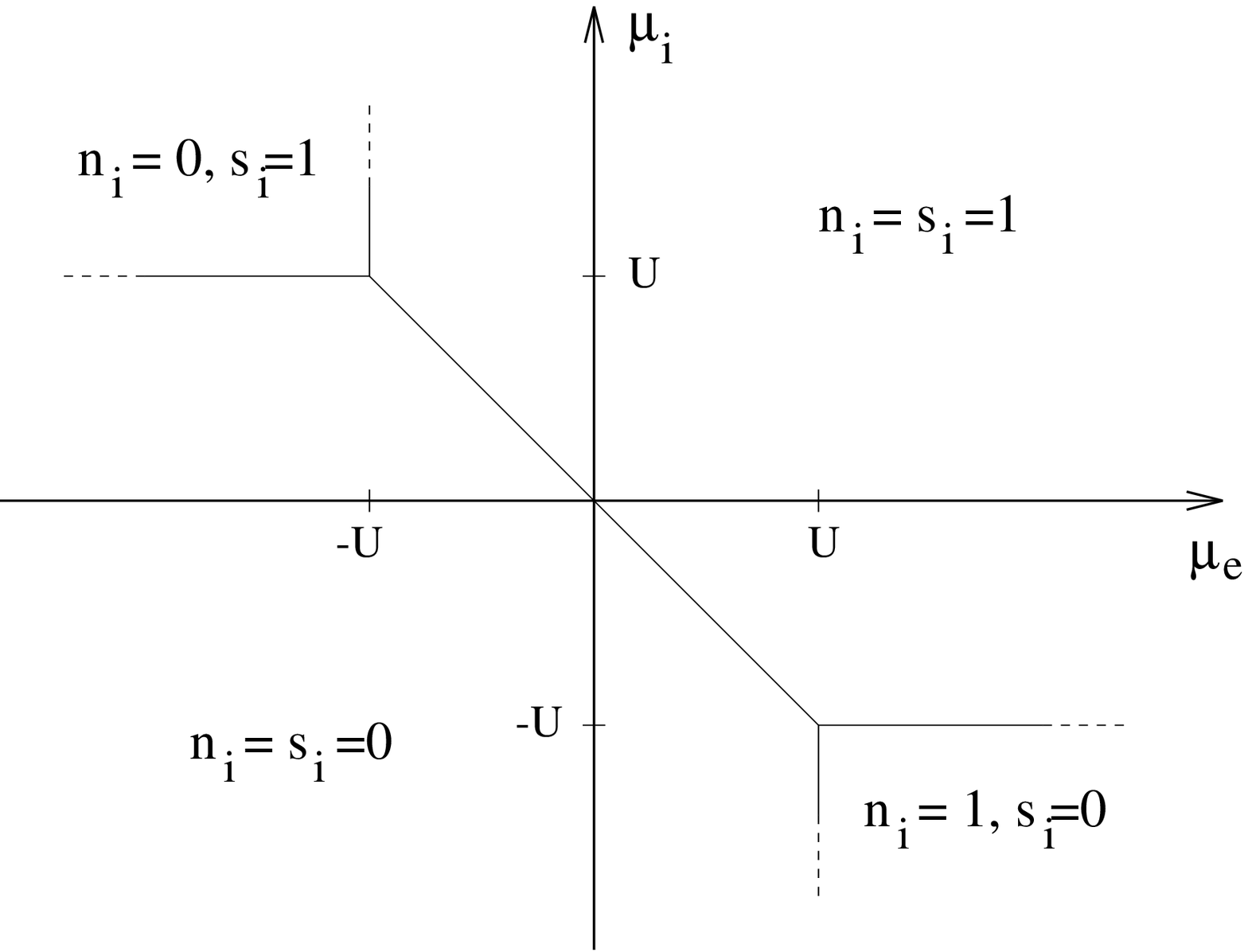,width=3.0in}}
\vskip 0.4cm
\caption{The phase diagram of the Falicov-Kimball model for $t=0$.}
\label{fig:pdt=0}

\vskip 0.5cm

\centerline{\psfig{figure=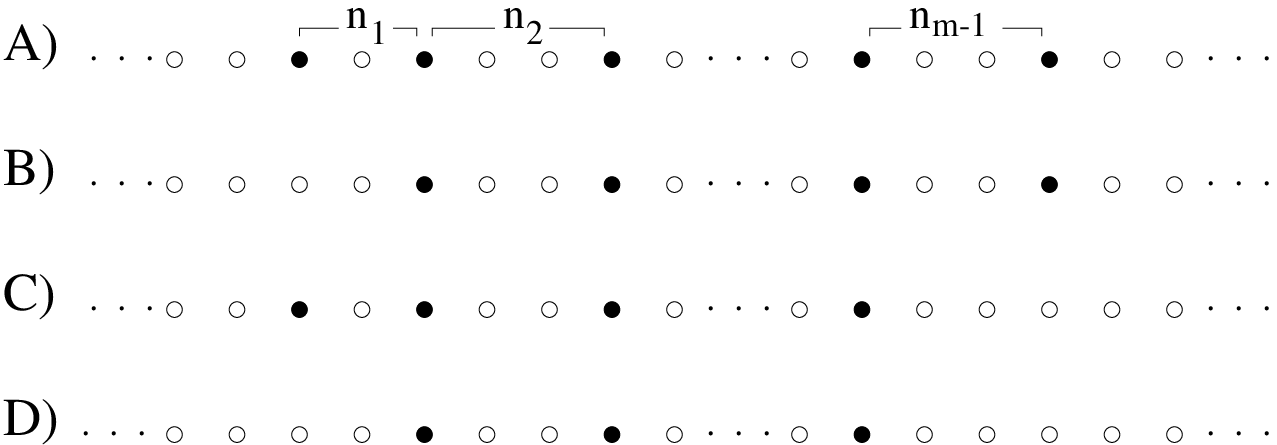,width=3.0in}}
\vskip 0.4cm
\caption{Ionic configurations needed to calculate the $m$-ion
interaction $V_m(n_1,n_2, ... n_{m-1})$. In $A$ there are $m$ ions
with successive separations $n_1, n_2, ... , n_{m-1}$. In $B$ the
left-most ion is
removed; in $C$ the right-most ion is removed; and in $D$ both the left-most
and right-most ions are missing.}
\label{fig:reco}

\newpage
\centerline{\psfig{figure=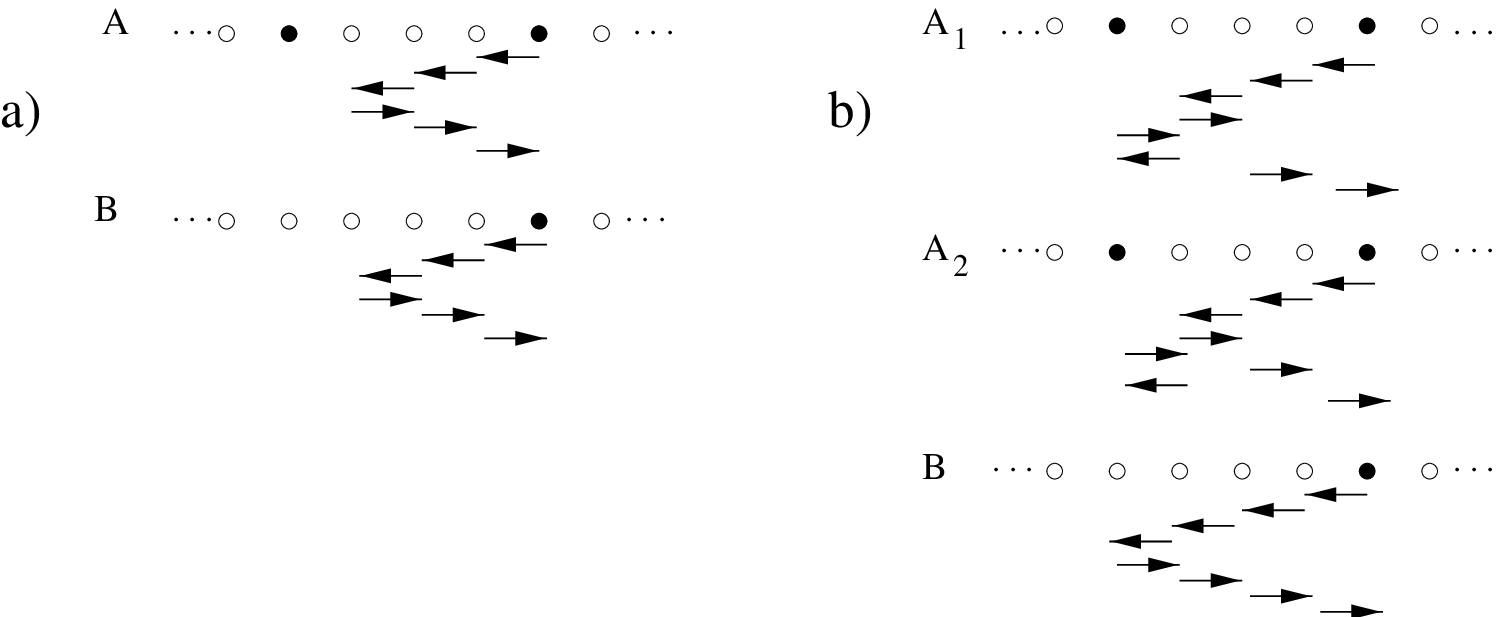,width=3.0in}}
\vskip 0.4cm
\caption{Examples of diagrams that, when the reconnection formula
(\protect{\ref{eqn:reco}}) is used, give an (a) zero (b) leading-order
contribution to the two-ion interaction $V_2(4)$. A full circle
represents an ion and an arrow denotes the hopping of an electron.
The contribution to the energy can depend on the order of the arrows
which determines the energy denominators, $p U$, that arise in
perturbation theory, where $p$ is the number of electrons away from
their ions. For example, in configuration $A_1$, six denominators are
$U$ and one is $2U$ whereas, in $A_2$, five denominators are $U$ and
two are $2U$.}
\label{fig:diag}
\end{figure}

The phase diagram for $t=0$ is shown in Fig.~\ref{fig:pdt=0}.
All the phase boundaries in the figure are multidegenerate in that any
phase obtained by mixing the two neighboring phases is
degenerate on the boundary. Our aim is to study systematically how this
multidegeneracy is lifted as $t/U$ increases from zero.

It is convenient to introduce the variables
\begin{eqnarray}
&& h \equiv(\mu_i + \mu_e)/2\ , \\
&& \Delta \equiv (\mu_i - \mu_e)/2 \ .
\end{eqnarray}

\noindent $U$ is assumed to be much larger than any physical parameter
in (\ref{eqn:ham}) and therefore $\Delta/U \ll 1$. This restriction on
$\Delta$ has the important consequence of fixing the total number of
electrons equal to the total number of ions, and  throughout the rest of
the paper, we will implicitly consider a neutral system,
$\sum_i n_i = \sum_i s_i$, where $n_i=a^\dagger_i a_i$.

When moving along the line $\mu_e = \mu_i$ in Fig.~\ref{fig:pdt=0} one notices
that, for negative values of $h$, the ground state corresponds to an
empty lattice ($n_i=s_i=0$). On the other hand, for $h$ positive
$n_i=s_i=1$. The point $h=0$ lies on the multi-degenerate
phase boundary where all phases associated with an arbitrary spacing
of the ions are degenerate.
To distinguish between the different degenerate states it is
convenient to introduce the labelling $\langle n_1, n_2,
.. n_m\rangle$ to denote a phase consisting of ions whose separations
(measured in lattice spacings) repeat periodically the sequence  $n_1,
n_2, .. , n_m$.  (Hence the phases
$n_i=s_i=1$ and $n_i=s_i=0$ can be
described as  $\langle 1 \rangle$ and $\langle \infty \rangle$ respectively.)

The multidegeneracy encountered on the phase boundaries of Fig.~\ref{fig:pdt=0}
is
due to the absence of interaction between the confined electrons. It
is natural to expect that, for $t/U \not= 0$, the hopping of electrons
will introduce an effective coupling between the ions, thus providing
a mechanism for the removal of the degeneracy. This intuitive picture can
be formalized using the defect-defect interactions introduced
by Fisher and Szpilka [\onlinecite{FS}]. In the present context, a defect
corresponds to an ion. Following [\onlinecite{FS}] the energy per
lattice site of phase $\langle n_1, n_2,..., n_m \rangle$ can be written as
$E_{\langle n_1, ... ,n_m \rangle} = E_{\rm tot}/ \sum_{i=1}^m n_i$, where
\begin{equation}
E_{\rm tot} =  m \sigma + \sum_{i=1}^m V_2(n_i) +\sum_{i=1}^m
V_3(n_i,n_{i+1})+...
\label{eqn:enexp}
\end{equation}
\noindent where $\sigma$ is the creation energy of an isolated
ion, $V_2(x)$ denotes the effective interaction between two
ions at a distance $x$, $V_3(x,y)$ the interaction of three ions with
spacings $x,\ y,$ and so on.
Although, for simplicity, we refer to the ion creation energy and
ion--ion interactions, it must be borne in mind that each ion is
associated with an electron.

When $t=0$ the electrons are confined to the ions.
 In this case $\sigma$ is readily
shown to be equal to $-2h$. For small $t/U$ we expect each electron to
be localized in a region around the associated ion. Using standard
perturbation theory one can obtain $\sigma$ to leading order in $t/U$
\begin{equation}
\sigma =-2 h - 2{ t^2/U} + {\cal O}(t^4/U^3) \ .
\end{equation}

\noindent The leading order corrections to $\sigma$ are associated with
virtual a process in which the electron hops to the site to the
immediate right (or left) of the ion and back again.

The general ion--ion interaction term, $V_m(n_1,n_2,...,n_{m-1})$, can
be obtained, at least in principle, through a reconnection formula
[\onlinecite{Bass}]. In terms of the four different ionic configurations
shown in Fig.~\ref{fig:reco}, this formula is
\begin{equation}
V_m(n_1, n_2, ... , n_{m-1}) = E_A - E_B - E_C + E_D\ .
\label{eqn:reco}
\end{equation}

In the absence of electron hopping, Eq.~(\ref{eqn:reco}) gives
$V_m=0$ for all values of $m$. Our aim is to calculate $V_m$ to
leading order in $t/U$. A simple way to
obtain the leading-order contribution to the interaction between two
ions occupying sites $0$ and $n$ relies on
perturbation diagrams in which the matrix element of $t\, a_i^\dagger
a_j$ is represented by an arrow from $j$ to $i$ and the ordering of
matrix elements follows the height on the page. First note that
diagrams involving disjoint sets of sites do not contribute to the
ground state energy.  In
addition,  diagrams involving fewer than $2n$ electron hops will not
contribute to $V_2(n)$ because, as illustrated in
Fig.~\ref{fig:diag}a, the contribution of every such diagram in
configuration A will be cancelled by a counter-diagram in
configuration B (or C).

It is then apparent that the leading-order contribution to $V_2(n)$
is due to diagrams where the sites $0$ and $n$ are just connected by $2n$
hoppings, as in Fig.~\ref{fig:diag}b, and is proportional
to  $t^{2n}/U^{2n-1}$. The
proportionality factor can be calculated by summing the contributions from all
relevant diagrams
\begin{equation}
V_2(n) = 2n { t^{2n}/U^{2n-1}} + {\cal O}( { t^{2n+2} / U^{2n+1}})\ .
\label{eqn:v2n}
\end{equation}

Fisher and Szpilka [\onlinecite{FS}] showed that, for systems where the
ion--ion interactions, $V_m$, decay sufficiently rapidly with the defect
spacings, a knowledge of the sign and convexity of $V_2(n)$ can
provide a considerable amount  of qualitative
information about the phase diagram of the system. Their analysis can
be applied in this context since the $V_m$ decay exponentially with
the spacings of the two outermost ions
(because the ion--ion interaction is mediated by the nearest-neighbor
hoppings
of electrons). Therefore, as a first approximation, we shall analyse
the phase diagram neglecting interactions that involve
more than two ions. Higher-order interactions will then be included
successively to resolve the finer details of the phase structure.

 In the two-ion interaction approximation the ground-state
configurations correspond to equispaced electron--ion pairs.
 Since $V_2(n)$ is always positive
and convex, as $h$ is varied from positive to negative,
${n}$ increases monotonically in steps of  one lattice spacing
[\onlinecite{FS}],
giving rise to the infinite sequence of phases
\begin{equation}
\langle 1 \rangle \to \langle 2 \rangle \to ... \to \langle \infty
\rangle \ .
\end{equation}

\noindent The phase $\langle n \rangle$ is stable over a
region of width
\begin{equation}
\Delta h_n \approx {n+1 \over 2} V_2(n-1) \approx (n^2 -1)\, {
t^{2n-2}/ U^{2n-3}} \ .
\end{equation}

The original multidegeneracy is not completely lifted by $V_2(n)$
because, on the
boundary between two phases, $\langle n \rangle$ and $\langle n+1
\rangle$, all mixed phases where the ions can be separated by
distances $n$ or $n+1$  are still degenerate. To
determine the finer structure of the phase diagram it is necessary to
consider the effect of higher-order ion interactions. These are not
easily obtained using the simple method outlined above since, when
there are more than two ions, it is
extremely difficult to keep track of the energy denominators
associated with the different orderings of the electron hoppings. However,
this problem can be circumvented by using Green's function
techniques. We first illustrate how this method can be used to
reproduce the result for $V_2(n)$.

To calculate $E_A$ in Eq.~(\ref{eqn:reco}) consider a system of
$n~+~1$ sites with ions at sites $0$ and $n$.
The single-particle energies are determined by the
eigenvalues of the x$(n~+~1)$-dimensional matrix, ${\bf M}$, where
\begin{equation}
M_{ij} = -U \delta_{i,j} (\delta_{i,0} + \delta_{i,n}) +t
(\delta_{i,j+1} \delta_{i,j-1})
\end{equation}
\noindent and the other matrix elements are zero.
Two of these energies occur near $-U$, and these are the ones we want to
sum over.  So we write
\begin{equation}
\label{CINT}
E_A = {1 \over 2 \pi i} \int_\Gamma {\rm Tr} \Biggl[ \Biggl(  z {\cal I} - {\bf
M}
\Biggr)^{-1} \Biggr] z \ dz\ ,
\end{equation}
where the contour $\Gamma$ encloses the region near $z=-U$ and ${\cal
I}$ is the identity matrix.  To evaluate the
trace we expand the matrix inverse in
Eq.~(\ref{CINT}) in powers of the $t$'s. Define a perturbation
$V_{ij} = t( \delta_{j,i+1} + \delta_{j,i-1} )$. Then
\begin{eqnarray}
& & \biggl( z {\cal I} - {\bf M} \biggr)_{ii}^{-1}
=  G_{ii} + G_{ii}V_{ij} G_{jj}V_{ji}G_{ii} + \nonumber \\
& & G_{ii}V_{ij} G_{jj}V_{jk}G_{kk}V_{kl}G_{ll}V_{li}G_{ii} + \dots
\end{eqnarray}
\noindent where $G_{ii}= [z-M_{ii}]^{-1}$.
Terms which are odd order in $t$ cannot contribute to the trace.

In this expansion one sees that, if $i$ is not an end site,
in order to involve all the $t$'s the matrix elements must
start at $i$, say, then increase to the highest number site
($n$), then decrease to the lowest number site ($0$) and
finally increase back to the original value $i$.  Alternatively,
the matrix elements could initially decrease. If $i=0$ or $n$ however, note
that the index can  only initially increase or decrease respectively.
So to leading order
\begin{equation}
\biggl( {z {\cal I} - {\bf M}} \biggr)_{ii}^{-1}
\approx C_i  G_{00} G_{ii} G_{n,n} \prod_{i=1}^{n-1} G_{ii}^2
\prod_{i=0}^{n-1} V_{i,i+1}^2 \ ,
\label{TRACE}
\end{equation}
where $C_i=1$ if $i=0$ or $i=n$ and $C_i=2$ otherwise.
The product over $G$'s does not include the end sites, because
these, in general, only appear once.  The starting site
appears an extra time and gives rise to the prefactor $G_{ii}$.
The term of order $t^{2n}$ in Eq.~(\ref{TRACE}) is
\begin{equation}
{\rm Tr} ({z {\cal I} - {\bf M}} )^{-1}
\approx t^{2n} \Biggl[ {2 \over (z+U)^3 z^{2n-2}} + {2n-2 \over
(z+U)^2 z^{2n-1}} \Biggr] \ .
\label{eqn:tr}
\end{equation}
Here the first term includes $C_1$ and $C_{n+1}$, both of which are unity.
The factor $2n-2$ comes from $\sum_{i=2}^n C_i$. Substituting
(\ref{eqn:tr}) in (\ref{CINT}) and calculating the integral using
residues gives (here and below we give the expressions only to leading order
in $t/U$)
\begin{eqnarray}
E_A&=& {1 \over 2 \pi i} \int_{\Gamma} t^{2n} \Biggl[ {2 \over
(z+U)^3 z^{2n-3} } + {2n-2 \over (z+U)^2 z^{2n-2}} \Biggr] dz
\nonumber \\
&=& (2n-2) {t^{2n} / U^{2n-1}}  \ .
\end{eqnarray}

\noindent Next, to use the reconnection formula (\ref{eqn:reco}), we need to
repeat the same calculation when one of the end ions is removed
(corresponding to configurations B and C in Fig.~\ref{fig:reco}). In
this case

\begin{eqnarray}
{\rm Tr} ( {\cal I} - {\bf M})^{-1}
= t^{2n} \Biggl[ {1 \over (z+U)^2 z^{2n-1} } + {2n-1 \over
(z+U) z^{2n}} \Biggr] \ .  \nonumber
\end{eqnarray}
Thus the perturbative contributions to $V_2$, denoted $E_B$ and $E_C$,  are
$E_B = E_C  =  -{t^{2n}/U^{2n-1}} $. Note than when both ions are
removed there are no longer any levels near $-U$. Hence $E_D=0$ and
use of the reconnection formula (\ref{eqn:reco}) gives $V_2(n) =2n\,
{t^{2n} / U^{2n-1}}$, in agreement with the expression (\ref{eqn:v2n}).

The method outlined above
can be extended to calculate the $m$-ion interaction $V_m$ for $m >
2$. As we shall show
below, $V_m(n_1, n_2, ... n_{m-1})$ depends, to leading order, only on
the separation of the two outermost ions in configuration $A$,
$n=\sum_{i=1}^{m-1} n_i$. The result is

\begin{equation}
V_m(n) = {(2n)!\over (2m-3)! (2n-2m+3)!} {t^{2n} \over  U^{2n-1}} \ .
\label{eqn:vm}
\end{equation}

\noindent To prove this consider Eq.~(\ref{TRACE}).  Note that
$m$ of the diagonal elements of $(z {\cal I} - M)^{-1}$ are
$(z+U)^{-1}$; the rest are $z^{-1}$.
If the initial $i$ corresponds to an ion, then a factor
$(z+U)^{2m-1}$ appears in the trace; otherwise the factor is  $(z+U)^{2m-2}$.
In the first case there are $m$ choices for $i$; two at the end with
$C_i=1$ and $m-2$ in the interior with $C_i=2$.  Thus
\begin{eqnarray}
{\rm Tr} \Biggl[ (  z {\cal I} - {\bf M} )^{-1}
\Biggr] &=& t^{2n} \Biggl\{ {(2m-2) \over (z+U)^{2m-1}
z^{2n-2m+2} } \nonumber \\
& & +{2n-2m+2 \over (z+U)^{2m-2} z^{2n-2m+3}} \Biggr\} \ .
\label{eqn:trzm}
\end{eqnarray}
Again we stress that the dependence of (\ref{eqn:trzm}) on the
position of the $m$ ions in the chain is only through $n$, the
distance between the two end defects.  Substituting in (\ref{CINT})
gives
\begin{eqnarray}
E_A & = & {(2n-2)! \over (2m-3)! (2n-2m+1)!}{ t^{2n} \over  U^{2n-1} } \ .
\end{eqnarray}
\noindent Similarly
\begin{eqnarray}
E_B &=& E_C  =  E_A (2m-3)/(2n-2m+2)  \\
E_D &=& E_B (2m-4) / (2n -2m +3) \ .
\end{eqnarray}

\noindent Finally the use of the reconnection formula (\ref{eqn:reco})
gives for the $m$-ion effective interaction, $V_m$, the result
(\ref{eqn:vm}). It should be pointed out that, in principle, the
leading order expression (\ref{eqn:vm}) could be dominated by
neglected terms of higher order in $t/U$ if $n$ is sufficiently large
(for fixed $t/U$) [\onlinecite{FS,Bass}]. However, Gruber {\em et
al.}\/ [\onlinecite{GLM}]
have shown that, for $m=2$, higher-order corrections to $V_2(n)$ are
dominated uniformly in $n$ by the expression (\ref{eqn:vm}), provided
that $t$ is replaced by $\tilde{t}= U [ \sqrt{U^2 + 4 t^2} -
U]/2t$. It seems plausible to expect that, upon renormalizing $t$ in
(\ref{eqn:vm}), their conclusion can also be extended to $m>2$.

We now consider how higher-order ion interactions modify the phase
diagram obtained in the
two-ion interaction approximation. Consider first $V_3$. This has the effect
of partially removing  the multidegeneracy on the $\langle n \rangle |
\langle n+1 \rangle$ boundaries by stabilizing the mixed phases
$\langle n, n+1 \rangle$. This happens because the energy difference
\begin{eqnarray}
& &  (2n+1)E_{\langle n, n+1 \rangle} - n E_{\langle n \rangle} -
(n+1)E_{\langle n+1 \rangle} =\nonumber \\
& &  V_3(n,n+1)+ V_3(n+1,n) - V_3(n,n) - V_3(n+1,n+1) \nonumber
\end{eqnarray}

\noindent is negative. The mixed phase $\langle n,n+1 \rangle$ has an
ion density, $2/(2n+1)$, intermediate between the pure phases $\langle
n \rangle$  and $\langle n+1 \rangle$.

The stability of the two new boundaries appearing at this stage of
approximation, namely $\langle n \rangle | \langle n,n+1 \rangle$ and
$\langle n,n+1 \rangle | \langle n+1 \rangle$ can be determined similarly by
considering four-ion interaction terms. Again they are unstable to the
appearance of the mixed phases $\langle n, n, n+1 \rangle$ and
$\langle n, n+1, n+1 \rangle$ respectively. Indeed, since all interaction
potentials are positive and decay exponentially with the
separation of the outmost ions, we can conclude that, at every stage
of the construction of the phase diagram, the introduction of
neglected higher-order interactions will lead to the stabilization of
mixed phases of increasingly long period.

To summarize: we have calculated the general $m$-ion interaction
potentials in the neutral Falicov-Kimball model to leading order in
$t/U$ at zero temperature.  We thereby iteratively construct
the ground-state phase diagram and conclude that the ion density
versus chemical potential, $h$, has the form of a complete devil's
staircase.

Extending the strategy for the iterative construction of the phase
diagram to more than one dimension is not trivial and is the focus of
an ongoing investigation.

JMY and CM acknowledge support from the EPSRC and ABH from the National
Science Foundation under grant 95-20175.  We thank
J. J\c{e}drzejewski, R. Lemansky, M. Rasetti and G. Watson for useful
discussions.

\end{document}